\documentclass{appolb}
\usepackage{graphicx}
\usepackage{xcolor}
\usepackage{amsmath,amssymb,amsfonts}
\usepackage{cite}
\usepackage{multirow}
\usepackage{framed}
\usepackage{slashed}
\usepackage{bbm}
\usepackage{array}
\usepackage{booktabs}
\usepackage{hyperref}
\usepackage{color}

\newcommand\sigmav {\langle\sigma v_\text{M{\o}l}\rangle}
\newcommand\rd {\ensuremath{\mathrm{d}}}
\newcommand\tW {\theta_{\mathrm{W}}}
\newcommand\ri {\ensuremath{\mathrm{i}}}

\begin{document}
\title{Sterile neutrino dark matter in a U(1) extension of the standard model\thanks{Presented by K. Seller at the on-line meeting {\it Matter to the Deepest}, Katowice, Poland, 15-17 September 2021.}
}
\author{
Károly Seller$^1$
\address{1. Institute for Theoretical Physics, ELTE E\"otv\"os Lor\'and University, \\ 
P\'azm\'any P\'eter s\'et\'any 1/A, H-1117 Budapest, Hungary}
}
\maketitle
\begin{abstract}
We explore the possibilities of dark matter production in a U(1) extension of the standard model, also called the super-weak model. The freeze-in and freeze-out mechanisms are described in detail, assuming the lightest sterile neutrino in the model as the dark matter candidate.
In both scenarios we present the favoured parameter space on the plane of super-weak coupling versus the new gauge boson mass. We discuss the experimental constraints limiting each case and outline possibilities of detection.
\end{abstract}


\section{Introduction}

The existence of dark matter is a well established experimental fact.
However, despite the extreme success of the standard model of particle physics in describing most of the observations in particle physics, 
it is unable to account for this seemingly invisible matter in the Universe.
Lots of models have been proposed to solve this riddle, mostly relying on either the modification of our understanding of gravity, or the extension of the standard model of particle physics.
A common pitfall of these theories is that they focus on a single aspect of beyond the standard model (BSM) physics, while leaving others unexplored. 
The super-weak model is an attempt at formulating an extension of the standard model \cite{Trocsanyi:2018bkm}
capable of explaining multiple sides of the BSM puzzle.
  
\section{The super-weak model}

The super-weak model is a simple extension of the standard model by a U(1)$_z$ gauge group, originally introduced in Ref.~\cite{Trocsanyi:2018bkm}. 
The model was designed to be phenomenologically simple, yet capable of explaining a number of shortcomings of the standard model. 
In particular, (i) the exploration of the origin of neutrino masses and neutrino oscillations have been started 
in Refs.~\cite{Iwamoto:2021wko,Karkkainen:2021syu}, (ii) cosmic inflation and electroweak vacuum-stability has been discussed in Ref.~\cite{Peli:2019vtp}, and (iii) dark matter scenarios with the lightest sterile neutrino have been studied in Ref.~\cite{Iwamoto:2021fup}.

The particle spectrum is extended to include three sterile neutrinos $N_i$, a singlet complex scalar $\chi$, and the gauge boson of the new U(1)$_z$ group $Z'$. The anomaly-free charge assignment and the spectrum is presented in Table~\ref{tab:ChargeAssignment}.

\begin{table}[t]
    \centering
    \begin{tabular}{cccc}\toprule
         & SU(2)$_\mathrm{L}$ & U(1)$_y$ & U(1)$_z$ \\
        \midrule
        $Q_\mathrm{L}$ & $\mathbf{2}$ & $1/6$ & $1/6$ \\
        $U_\mathrm{R}$ & $\mathbf{1}$ & $2/3$ & $7/6$ \\
        $D_\mathrm{R}$ & $\mathbf{1}$ & $-1/3$ & $-5/6$ \\
        $L_\mathrm{L}$ & $\mathbf{2}$ & $-1/2$ & $-1/2$ \\
        $N_\mathrm{R}$ & $\mathbf{1}$ & 0 & $1/2$ \\
        $e_\mathrm{R}$ & $\mathbf{1}$ & $-1$ & $-3/2$ \\
        \midrule
        $\phi$ & $\mathbf{2}$ & $1/2$ & 1 \\
        $\chi$ & $\mathbf{1}$ & 0 & $-1$ \\
        \bottomrule
    \end{tabular}
    \caption{Particle content and charge assignment of the super-weak model.}
    \label{tab:ChargeAssignment}
\end{table}


\subsection{The gauge sector}

The super-weak gauge group gets spontaneously broken when the scalars in the theory acquire non-zero vacuum expectation values. Similarly to the standard model, only the electromagnetic U(1)$_\mathrm{em.}$ gauge group remains a true symmetry of the theory,
\begin{equation}
    \mathrm{G}_\mathrm{SW}=\mathrm{SU}(3)_\mathrm{c}\otimes\mathrm{SU(2)_\mathrm{L}\otimes}\mathrm{U}(1)_y\otimes\mathrm{U}(1)_z\to \mathrm{SU}(3)_\mathrm{c}\otimes\mathrm{U(1)}_\mathrm{em}
\end{equation}
Through the Higgs-mechanism some of the gauge bosons obtain masses. The gauge eigenstates $(B_\mu,~W^3_\mu,~B'_\mu)$ are rotated to the mass eigenstates $(A_\mu,~Z_\mu,~Z'_\mu)$, where in contrast to the standard model,
we can define three angles of rotation: $\theta_\mathrm{W}$ (standard model Weinberg angle), $\theta_Z$ and $\theta_\epsilon$. The rotation by $\theta_\epsilon$ (mixing in the $B_\mu-B'_\mu$ plane) is unphysical as it gets cancelled in the Lagrangian by the requirement of the photon being massless.

In the super-weak model the standard model weak neutral current is modified due to the small, but non-zero mixing between the $Z$ and $Z'$ bosons ($\theta_Z\ll 1$). The covariant derivative for the neutral currents is
\begin{equation}
    \mathcal{D}_\mu^\mathrm{neut.}\supset -\ri(\mathcal{Q}_AA_\mu +\mathcal{Q}_ZZ_\mu+\mathcal{Q}_{Z'}Z'_\mu)\,,
\end{equation}
where the effective charges are defined by
\begin{subequations}
\begin{align}
    &\mathcal{Q}_A = (T_3+y)|e|\equiv\mathcal{Q}^\mathrm{SM}_A\,, \\
    &\mathcal{Q}_Z = (T_3\cos^2\tW - y\sin^2\tW)g_{Z^0}\cos\theta_Z-\zeta g_z\sin\theta_z = \mathcal{Q}_Z^\mathrm{SM} + \mathcal{O}\left(\frac{g_z^2}{g_{Z^0}^2}\right)\,, \\
    &\mathcal{Q}_{Z'}= (T_3\cos^2\tW - y\sin^2\tW)g_{Z^0}\sin\theta_Z-\zeta g_z\cos\theta_z \,.
\end{align}
\end{subequations}
Here we introduced the notation $g_{Z^0}^2 = g_ \mathrm{L}^2 + g_y^2$ and the effective U(1)$_z$ charge $\zeta = z-\eta y$ (charges depend on the fields), where $\eta$ is connected to the gauge kinetic mixing of the U(1) groups.


\subsection{Parameters of the model}

In total there are 5 parameters in the model that are relevant for dark matter studies: (i) the gauge coupling $g_z$, (ii) the mass of the $Z'$ boson, (iii) the $Z-Z'$ mixing angle $\theta_Z$, (iv) the U(1) gauge mixing $\eta$, and (v) the smallest of the sterile neutrino masses. The dark matter candidate in the model is the lightest sterile neutrino $N_1$; the other two are assumed to be heavy, and decouple from the dark matter calculations.
\begin{enumerate}
\itemsep=-2pt
    \item[(i)] The gauge coupling $g_z$ has to be small when compared to the electroweak couplings, $\mathcal{O}(g_z/g_{Z^0})\ll 1$ in order to 
    obey various particle physics constraints, such as electroweak precision measurements.
    \item[(ii)] The mass of the $Z'$ boson $M_{Z'}$ is chosen such that $M_{Z'}\ll M_Z$ is satisfied. In particular, we will use $M_{Z'}=[10,200]~$MeV, where the upper bound is needed to avoid direct decays into muons.
    \item[(iii)] Assuming that (i) and (ii) are satisfied, the $Z-Z'$ mixing angle is
    \begin{equation}
        \tan(2\theta_Z)=\frac{4\zeta_\phi g_z}{g_{Z^0}}+\mathcal{O}\left(\frac{g_z^2}{g^2_{Z^0}}\right)\ll 1\,.
    \end{equation}
    As such the mixing $\theta_Z$ is not a free parameter, but similarly to the Weinberg angle, it is a function of the couplings.
    \item[(iv)] The value of the U(1) gauge mixing parameter $\eta$ depends on the renormalization scale $\mu$, and its exact value can be obtained from its renormalization group equation. At relevant scales $\eta=\mathcal{O}(0.1)$. To simplify calculations, we will use $\eta=0$, since it does not introduce any qualitative difference in the results.
    \item[(v)] We assume that the lightest sterile neutrino is either a keV-scale (freeze-in), or an MeV-scale (freeze-out) particle. Meanwhile for the heavier sterile neutrinos $M_{2,3}\gtrsim M_Z$, which allows for active neutrino masses in their allowed region with similar Yukawa couplings as in the charged lepton sector.
\end{enumerate}


\section{Dark matter production}

Dark matter production is most commonly achieved in extensions of the standard model by using so-called portals: 
interactions connecting the dark sector to the standard model. In the super-weak model we consider the so-called vector boson portal, with $Z'$ being the main connection between standard model particles and the dark matter candidate sterile neutrino $N_1$. 

The $Z'$ gauge boson is assumed to be light, so that it can decay only to (i) an electron-positron pair, (ii) a pair of standard model neutrinos, or (iii) a pair of $N_1$ sterile neutrinos (provided that $2M_1<M_{Z'}$). These vertices (up to small corrections of order $\mathcal{O}(g_z^2/g_{Z^0}^2)$) are given as
\begin{equation}
\label{eq:vertices}
    \Gamma^\mu_{Z'\nu_i\nu_i}\simeq\Gamma^\mu_{Z'N_1N_1}\simeq -\ri\frac{g_z}{2}\gamma^\mu\,, \quad
    \Gamma^\mu_{Z'ee}\simeq \ri g_z\gamma^\mu\left[2\cos^2\tW-\frac{1}{2}\right]\,.
\end{equation}
Vertices with heavier fermions may appear in scattering processes, however the mass of $Z'$ sets the energy scale where dark matter is dominantly produced, as such the main contribution will come from only those listed in Eq.~\eqref{eq:vertices}.

In cosmology, we use the Boltzmann equation
to describe the evolution of a particle species exposed to some interactions in an expanding Universe filled with a finite temperature plasma of particles. It is convenient to define the comoving number density $\mathcal{Y}=n/s$ to factor out Hubble expansion. The Boltzmann equation for a particle species $a$ produced via scatterings and decays is
\begin{subequations}
\label{eq:Boltzmann}
\begin{align}
    \frac{\rd\mathcal{Y}_a}{\rd z} &=\sqrt{\frac{\pi}{45}}g^*(T)\frac{m_\mathrm{Pl}\Lambda}{z^2}\left\langle\sigma_{ab\to f_1f_2}v_\text{M{\o}l} \right\rangle\left[\frac{\mathcal{Y}_a^\mathrm{eq}\mathcal{Y}_b^\mathrm{eq}}{\mathcal{Y}_{f_1}^\mathrm{eq}\mathcal{Y}_{f_2}^\mathrm{eq}}\mathcal{Y}_{f_1}\mathcal{Y}_{f_2}-\mathcal{Y}_a\mathcal{Y}_b\right]\\
    &+\sqrt{\frac{45}{4\pi^3}}\frac{g^*(T)}{g_s^*(T)}\frac{m_\mathrm{Pl}z}{\Lambda^2}\langle\Gamma_{a\to f'_1f'_2}\rangle \left[
    \frac{\mathcal{Y}_{f'_1}\mathcal{Y}_{f'_2}}{\mathcal{Y}_{f'_1}^\mathrm{eq}\mathcal{Y}_{f'_2}^\mathrm{eq}}\mathcal{Y}_{a}^\mathrm{eq}-\mathcal{Y}_a
    \right]\,.
\end{align}
\end{subequations}
Here $z=\Lambda/T$ with $\Lambda$ being a relevant energy scale in the problem, while 
the notation $\langle\dots\rangle$ denotes
thermal averaging. 

The thermally averaged cross section is given by
\begin{equation}
    \sigmav = \frac{1}{8m_\mathrm{in}^2T[K_2(m_\mathrm{in}/T)]^2}\int_{4\mu^2}^\infty\rd s~\sigma(s)(s-4m_\mathrm{in})^2\sqrt{s}K_1\left(\frac{\sqrt{s}}{T}\right)
\end{equation}
where $\mu=\max(m_\mathrm{in},m_\mathrm{out})$, and $K_i(x)$ are the modified Bessel functions of the second kind.
Notice that (i) the integral can be dominated by the resonance in $\sigma(s)$, and (ii) at low temperatures $\sigmav(T\ll m_\mathrm{in})\to 0$, which can lead to decoupling.

In contrast to scatterings, thermal averaging for the decay rate of a particle with mass $m$ is known analytically,
\begin{equation}
    \langle\Gamma\rangle = \frac{K_1(m/T)}{K_2(m/T)}\Gamma\,.
\end{equation}
The thermally averaged decay rate is a monotone increasing function of time, with $\langle\Gamma\rangle(T\ll m)\to\Gamma$.


\subsection{Freeze-out scenario}

The freeze-out mechanism of dark matter production assumes that the dark sector was in equilibrium with the cosmic plasma at high temperatures. 
As the temperature falls, processes which involve the creation and annihilation of dark matter particles become slow with respect to the Hubble rate, and a non-zero abundance can freeze-out.
The mechanism thus suggests that the relevant processes to consider are annihilations of standard model particles producing our dark matter candidates, the $N_1$ sterile neutrinos.

In the super-weak model, for the freeze-out scenario we consider the lightest sterile neutrinos to have masses of $\mathcal{O}(10)~$MeV, thus freeze-out will happen at $\mathcal{O}(1)~$MeV temperatures.
Assuming that the active-sterile mixing is tiny, there are only two processes which contribute to freeze-out, the annihilation of electrons or standard model neutrinos into sterile neutrinos.

\begin{figure}
    \centering
    \includegraphics[width=0.7\linewidth]{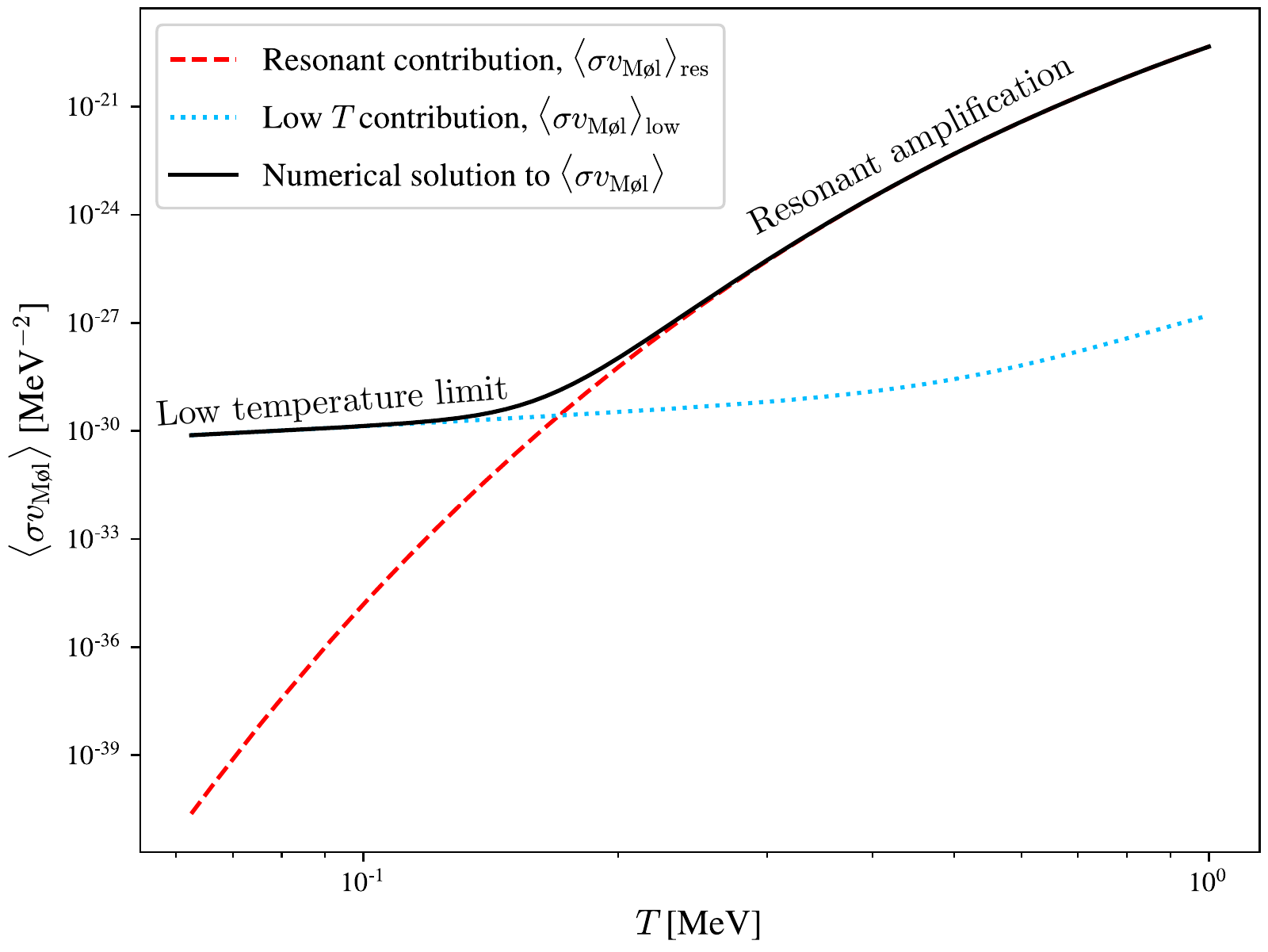}
    \caption{The thermally averaged cross section for $M_1=10~$MeV and $M_{Z'}=30~$MeV. At higher temperatures $\sigmav$ is totally dominated by the resonance contribution.}
    \label{fig:TACS}
\end{figure}

In the freeze-out mechanism, the smaller the coupling, the larger the relic density. 
As large couplings are constrained by experiments (e.g. NA64) this usually leads to overproduction of dark matter.
This well known issue can be circumvented by exploiting resonant production of dark matter: having the mediator $Z'$ be close to twice the mass of $N_1$. 
In Fig.~\ref{fig:TACS} we showcase the resonant amplification of the thermally averaged cross section.
\begin{figure}
    \centering
    \includegraphics[width=0.85\linewidth]{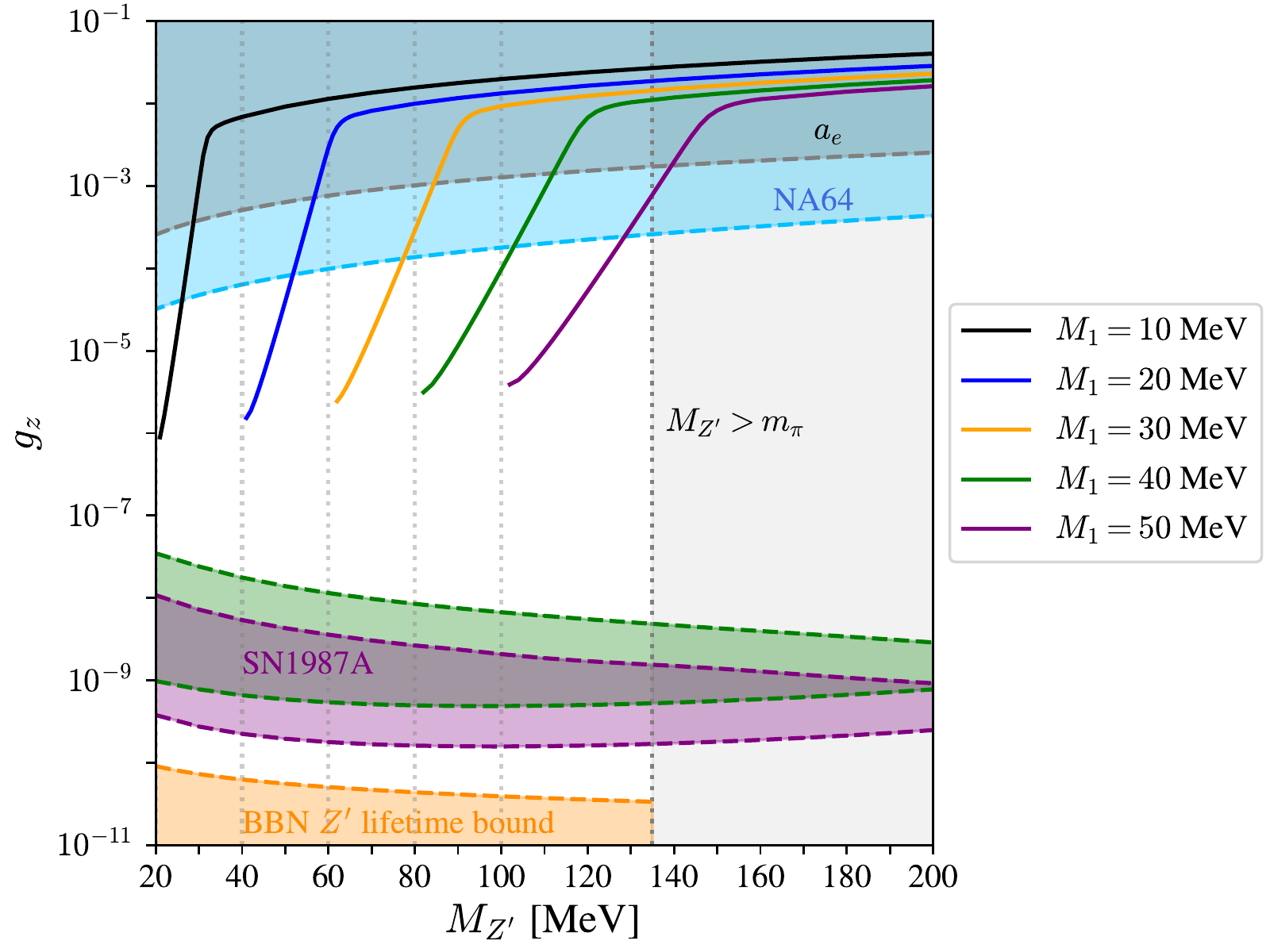}
    \caption{The parameter space in the freeze-out scenario: required couplings $g_z$ versus the mass of the new gauge boson $M_{Z'}$ for various sterile neutrino masses $M_1$. Constraints are detailed in Sec.~\ref{sec:ExpC}}
    \label{fig:FOPS}
\end{figure}

We have solved the Boltzmann equation Eq.~\eqref{eq:Boltzmann} for the $N_1$ sterile neutrino in the freeze-out scenario, and tuned the parameters of the model ($M_1$, $M_{Z'}$, and $g_z$) to reproduce the final relic densities required for the measured dark matter energy density.
We show the parameter space with various constraints in Fig.~\ref{fig:FOPS}.
We found that resonant amplification of the thermally averaged cross section is mandatory in order to avoid
established upper bounds on the couplings.
In the super-weak model, there is a non-vanishing parameter space, where the lightest sterile neutrinos can account for the total dark matter energy density.
The effect of the resonance is seen in the figure for $M_{Z'}\simeq 2M_1$, indeed without this amplification of the thermally averaged cross section the parameter space would be excluded by NA64.


\subsection{Freeze-in scenario}

Contrary to the freeze-out mechanism in the freeze-in case, the dark sector is never in equilibrium with the rest of the cosmic plasma. 
In particular it may be assumed that the dark particle densities are zero at some high temperatures (for example at $T_\mathrm{rh}$, after reheating). 
After the zero initial abundance, dark matter is created slowly by decays of heavier particles.
The mechanism requires that dark matter remains suppressed compared to equilibrium densities, which forces us to choose feeble couplings of $g_z=\mathcal{O}(10^{-10})$ or less. 
Since the dark matter particles are assumed to be stable, once the heavier particles have all decayed, a constant dark matter abundance freezes in.
The mechanism suggests that the important processes are decays of heavier particles into dark matter.

An advantage of the freeze-in mechanism could be that it has more parameters. The freeze-out scenario had only parameters connected to the particle physics model (coupling and masses), meaning that cosmology was essentially fixed. 
In the freeze-in case however, we have the freedom of choosing initial values: the initial temperature $T_0$ and the comoving density at that time $\mathcal{Y}_i(T_0)$.
If we choose $T_0$ to be the reheating temperature, then these initial conditions have a deep connection to the underlying cosmological model, which the freeze-out case did not have.
A detailed analysis of the choice of initial conditions shows that their exact values do not really affect low-energy results: if $T_0\ll \Lambda$ and $\mathcal{Y}_i(T_0)\ll\mathcal{Y}_\infty$, then any value of these 2 parameters will result in the same dark matter relic abundance. 
Here $\Lambda$ is the temperature (energy) scale where dark matter is dominantly produced, i.e., the mass of the decaying particle producing dark matter.
Choosing not to comply with these restrictions on the initial conditions will lead to interesting results, nevertheless these will usually require fine-tuning to fit to measurements, and for simplicity we avoid it in this work (we use $\mathcal{Y}_i(T_0)=0$).

In the super-weak model, for the freeze-in scenario we use lighter sterile neutrinos with masses of $\mathcal{O}(10)~$keV. 
The requirement for a feeble coupling tells us that the only relevant way to produce dark matter will be through the decays of $Z'$ bosons. 
However, as $Z'$ is also a dark sector particle we have no reason to assume that their initial abundance was different from that of the sterile neutrinos. 
This means that we have to solve the Boltzmann equation for the out-of-equilibrium distributions of $Z'$ bosons as well of the $N_1$ sterile neutrinos. 
The two form a coupled system of differential equations, where $Z'$ is mainly created by the coalescence of standard model leptons, $e^-+e^+\to Z'$ and $\nu_i+\nu_i\to Z'$, and destroyed by the inverse of these processes along with the decay $Z'\to N_1+N_1$ (the coalescence of $N_1$ neutrinos will be negligible due to the small population of $N_1$ particles).
In fact, since $N_1$ is only created by $Z'$ decays, one can approximate the final sterile neutrino density based on the branching ratio of $Z'$ into sterile neutrinos as $\mathcal{Y}_\infty=\max(\mathcal{Y}_{Z'})\mathcal{B}(Z'\to N_1+N_1)$.

\begin{figure}
    \centering
    \includegraphics[width=0.8\linewidth]{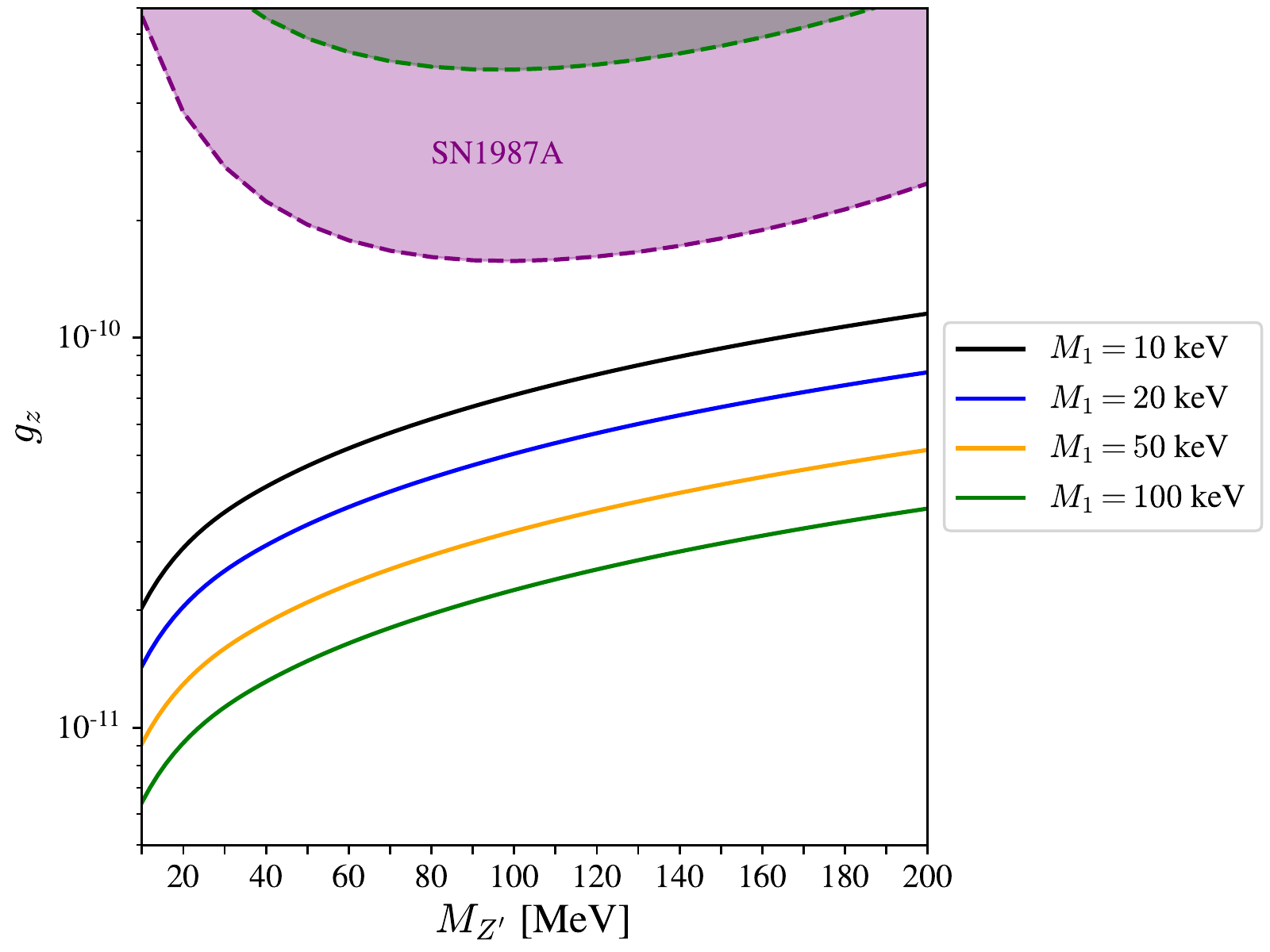}
    \caption{Parameter space in the freeze-in scenario: the required couplings to reproduce dark matter abundances are plotted versus the mass of the mediator $Z'$ boson for various sterile neutrino masses $M_1$. Constraints can be obtained on the model from astrophysical observations, most notably from the supernova SN1987A.
    }
    \label{fig:FreezeIn}
\end{figure}

We show the favoured parameter space we obtained for the freeze-in scenario in the super-weak model in Fig.~\ref{fig:FreezeIn}. 
As expected, the couplings are required to be feeble, $g_z\lesssim10^{-10}$.
Additionally, we can observe that for larger sterile neutrino masses, the required couplings decrease: this is due to the fact that $0.265=\Omega_\mathrm{DM}\propto M_1\mathcal{Y}_\infty\propto g_z^2M_1$ (this scaling is essentially exact due to the sterile neutrinos being very light, $M_1\ll M_{Z'}$, and the relic abundance being independent of $M_1$).

\subsection{Experimental constraints}
\label{sec:ExpC}

In this section we review the experimental constraints that apply to our model, both in the freeze-out and in the freeze-in scenario.
We detail those that have already been checked (such as those shown in Figs.~\ref{fig:FOPS}-\ref{fig:FreezeIn}), and outline our future work in this field.

The {\em freeze-out mechanism is mainly constrained by particle physics experiments}: in particular, the measurement of the anomalous magnetic moment of the electron, and NA64. 
These experiments provide stringent upper bounds on the coupling as a function of the $Z'$ mass, as the shaded areas indicate in Fig.~\ref{fig:FOPS}. 
The stronger constraint arises from NA64, which looks for missing energy in semi-bremsstrahlung processes, where in the final state instead of a photon, a dark photon is created.
This dark photon can decay into invisible particles (such as sterile particles, or even the standard model neutrinos), which would produce an event with missing measured energy.
The non-observation of such events provides an upper bound for the dark photon model (in particular it constrains the kinetic mixing) which can be translated to our model. In Fig.~\ref{fig:FOPS} we see that NA64 upper bound makes the use of resonant dark matter production necessary.
Astrophysical constraints provide only a weak lower bound on the parameter space which is avoided in our model.

Next to these upper bounds provided by particle physics, we also find a tentative cosmology-based upper bound on the $Z'$ mass, $M_{Z'}\leq m_\pi$.
This bound is based on the pion-enhanced proton-to-neutron conversion before the Big Bang nucleosynthesis (BBN).
If the $Z'$ bosons are allowed to decay to pions, then those pions can affect the ratio of protons to neutrons at the onset of BBN, thus modifying standard nucleosynthesis.
This provides a bound on the pion production rate.
The calculation of this rate is complicated, and requires further work.
However, it is not strictly necessary in our model, since the favoured parameter region where $M_{Z'}>m_\pi$ is already excluded by e.g.~NA64 (as shown in Fig.~\ref{fig:FOPS}).

{\em Astrophysics and cosmology based bounds become more relevant for the freeze-in scenario.}
In this case the feeble coupling $g_z\lesssim 10^{-10}$ makes it impossible to measure the effects of the new interaction in current particle physics experiments.
However cosmology and certain astrophysical objects may provide us with meaningful constraints which operate exactly in this feeble coupling range.
Firstly, stellar cooling is an established way of constraining models with light new particles (e.g. certain models with axions).
Our dark particles are much heavier than those investigated in stellar cooling reports, and these constraints are avoided.
Secondly, supernova explosions can be used to constrain BSM models based on energy loss in terms of invisible particles, or the production of gamma rays.
These bounds are not trivially avoided in our model. In fact they are relevant.
The supernova cooling bound is based on the single experimentally measured
explosion of SN1987A.
The phenomenological bound says that the luminosity of dark particles cannot be higher than
that of the standard model neutrinos.
Due to the opacity effects within the supernova, this constraint provides a exclusion band in the parameter space, such as the ones shown in Fig.~\ref{fig:FOPS}-\ref{fig:FreezeIn}. 
In both cases the calculated exclusion bands are approximations: the green one only takes into account interactions with the muons, whereas the purple one also includes the process $e^-+e^+\to Z'$ as well.
Of course, many more processes are available in our model, and their inclusion within this calculation is an ongoing project.

In conclusion, the freeze-out scenario is mostly constrained by particle physics, while freeze-in is mostly constrained by astrophysics.
In both cases we see the bounds are near the parameter space we are interested in.
On the particle physics side, future experiments, such as e.g., Belle II, NA64, or LDMX will be able to probe the coupling range $g_z\in[10^{-6}-10^{-4}]$, and provide us with the necessary data to test our model.
On the astrophysics side, we can only hope for a new supernova explosion event not too far from Earth. 
Such a measurement would allow us to increase our understanding of these extraordinary events, and to provide us with luminosity measurements of much higher accuracy.


\section{Conclusions}

In this talk, we have shown that the super-weak model can account for dark matter in the form of the lightest sterile neutrino. 
The freeze-out and freeze-in scenarios were investigated and shown to be able to provide us with a non-vanishing parameter space which avoids current experimental constraints.
Some of these constraints were briefly explored, and the possibility of future experiments to test our model was established.


\providecommand{\href}[2]{#2}\begingroup\raggedright\endgroup

\end{document}